\documentclass[journal=jacsat,manuscript=article]{achemso}

\usepackage[version=3]{mhchem} 

\author{Qingchen Yuan}
\altaffiliation{Contributed equally to the work.}
\affiliation{MOE Key Laboratory of Material Physics and Chemistry under Extraordinary Conditions, and Shaanxi Key Laboratory of Optical Information Technology, School of Science, Northwestern Polytechnical University, Xi'an 710072, China}

\author{Liang Fang}
\altaffiliation{Contributed equally to the work.}
\affiliation{MOE Key Laboratory of Material Physics and Chemistry under Extraordinary Conditions, and Shaanxi Key Laboratory of Optical Information Technology, School of Science, Northwestern Polytechnical University, Xi'an 710072, China}
\author{He Yang}
\altaffiliation{Contributed equally to the work.}
\affiliation{Department of Electronics and Nanoengineering, Aalto University, Espoo, FI-00076, Finland}
\author{Xuetao Gan}
\email{xuetaogan@nwpu.edu.cn}
\affiliation{MOE Key Laboratory of Material Physics and Chemistry under Extraordinary Conditions, and Shaanxi Key Laboratory of Optical Information Technology, School of Science, Northwestern Polytechnical University, Xi'an 710072, China}
\author{Vladislav Khayrudinov}
\affiliation{Department of Electronics and Nanoengineering, Aalto University, Espoo, FI-00076, Finland}
\author{Harri Lipsanen}
\affiliation{Department of Electronics and Nanoengineering, Aalto University, Espoo, FI-00076, Finland}
\author{Zhipei Sun}
\affiliation{Department of Electronics and Nanoengineering, Aalto University, Espoo, FI-00076, Finland}
\alsoaffiliation{QTF Centre of Excellence, Department of Applied Physics, Aalto University, Espoo, FI-00076, Finland}
\author{Jianlin Zhao}
\email{jlzhao@nwpu.edu.cn}
\affiliation{MOE Key Laboratory of Material Physics and Chemistry under Extraordinary Conditions, and Shaanxi Key Laboratory of Optical Information Technology, School of Science, Northwestern Polytechnical University, Xi'an 710072, China}

\title[An \textsf{achemso} demo]
  { Low-power Continuous-wave Second Harmonic Generation in Semiconductor Nanowires}

\abbreviations{IR,NMR,UV}
\keywords{Semiconductor nanowires, second harmonic generation, photonic crystal cavity}

\begin{document}
\begin{abstract}
  Semiconductor nanowires  (NWs)  are promising for realizing various on-chip nonlinear optical devices, due to their nanoscale lateral confinement and strong light-matter interaction. However, high-intensity pulsed pump lasers are typically needed to exploit their optical nonlinearity because light couples poorly with nanometric-size wires. Here, we demonstrate microwatts continuous-wave light pumped second harmonic generation (SHG) in AlGaAs NWs by integrating them with silicon planar photonic crystal cavities. Light-NW coupling is enhanced effectively by the extremely localized cavity mode at the subwavelength scale.  Strong SHG is obtained even with a  continuous-wave laser excitation with a pump power down to $\sim$3 $\mu$W, and the cavity-enhancement factor is estimated around 150. Additionally, in the integrated device, the NW's SHG is more than two-order of magnitude stronger than third harmonic generations in the silicon slab, though the NW only couple s with less than 1\% of the cavity mode. This significantly reduced power-requirement of NW's nonlinear frequency conversion would promote NW-based building blocks for nonlinear optics, specially in chip-integrated coherent light sources, entangled photon-pairs and signal processing devices. 
\end{abstract}

\section{Introduction}
 Semiconductor nanowires (NWs) represent nanoscale building blocks for photonic and optoelectronic devices, including low-threshold lasers \cite{Agarwal}, high-performance photodetectors \cite{Hayden,Fang}, and passive waveguides \cite{Piccione}. Besides these attributes and functions endowed by quantum-confined carrier dynamics, semiconductor NWs with no inversion symmetry also possess attractive second-order optical nonlinearities, such as ZnO~\cite{Johnson}, ZnTe~\cite{Liu}, ZnSe~\cite{Barzda}, ZnS~\cite{Hu}, CdS~\cite{Yu}, CdTe~\cite{Xin}, GaAs~\cite{Chen1}, GaN~\cite{Long}, etc. Benefited from their nanoscale diameters and high refractive indices, semiconductor NWs have higher nonlinear coefficient $\chi^{(2)}$ than their bulk counterpart for the effect of Lorentz local field~\cite{Zhuo,Sanatinia}. Further, their mechanical flexibility allows effective modulations of the tensor of $\chi^{(2)}$ by external bending and/or twisting strains~\cite{Han}. Because of NWs' high surface area to volume ratios, their $\chi^{(2)}$ could be modified as well using surface decoration or doping~\cite{Jassim}. NWs' intriguing second-order nonlinearities enable wide implementations of second harmonic generation (SHG) for nanoscale optical correlator~\cite{Yu}, crystallographic mapping~{ \cite{Neeman,Ren,Pimenta,Timofeeva},  defect identification~\cite{Liu2}, and nonlinear optical microscopy~\cite{Nakayama}.
	
However, previous studies involving bare NWs all present substantially low SHG conversion efficiency ($\sim10^{-6}$)~\cite{Hu,Ren} owing to their intrinsically small cross-section and poor spatial overlap with the light field. It therefore requires ultrahigh excitation power to produce moderately strong SHG in NWs that could be utilized for practical applications. Until now, the reported SHG experiments of NWs were realized using pulsed lasers with high peak powers ($\sim100$ W)~\cite{Hu,Ren}, which limits their device applications, specially for nanoscale coherent light sources and integrated photonic circuits. To improve SHG efficiencies in NWs, hybrid nanostructures consisted of NWs and metal materials were widely exploited to provide localized surface plasmons for light field enhancement~\cite{Grinblat,Casadei,Richter}. Unfortunately, the large ohmic loss and low coupling efficiency of the plasmonic nanostructures still make it inevitable to pump SHGs with pulsed lasers. Guiding the pump light along a NW could extend light-matter interaction length and has potential to improve SHG efficiency~\cite{Sergeyev}. However, phase matching and efficient waveguide-coupling are still challenging in the nanoscale footprint. 
	
Here, we demonstrate that the SHG from an AlGaAs NW could be significantly enhanced with a planar photonic crystal (PPC) cavity. As one of the optical cavities with the highest $Q/V_{mode}$ factor, resonant modes in PPC cavities have extremely high density of photons, which have been practically employed to realize strong photon-atom coupling and near-thresholdless lasers~\cite{Englund,Prieto}. Here, $Q$ and $V_{mode}$ are quality factor and mode volume of the cavity mode. In PPC cavities, the modes' transverse confinements are provided by the photonic bandgap of the periodic air-holes, allowing for the design of nanoscale cavities with high $Q$ factors. NWs could therefore effectively overlap with cavity's nanoscale mode. In the vertical direction, the modes are confined by the total internal reflection of the subwavelength thick silicon slab, presenting strong evanescent fields at the slab surface to couple with the integrated NW. Combining with cavity's high density of electrical field and its effective mode-coupling with NWs, the NW-PPC cavity presents a reliable configuration to enhance light-matter interactions in NW. For instance, the first NW laser at the telecom wavelength was realized by integrating an InP NW onto a silicon PPC cavity~\cite{Yokoo}. In this letter, continuous-wave (CW) light pumped SHG in NWs is reported, which is assisted by sufficient light-matter interactions in NW-PPC cavities. The excitation power is remarkably reduced down to a level of few microwatts. The cavity-enhancement factors are estimated around 150. 
\section{Device fabrication}

Figure~\ref{<device>}(a) schematically displays the precise alignment of a NW with a PPC cavity as well as the operation of cavity-enhanced SHG. The PPC cavities are fabricated in a silicon-on-insulator wafer with a 220 nm thick top silicon layer. The PPC patterns are defined using electron beam lithography, which are then transferred into the top silicon layer with the inductively coupled plasma dry etching. To ensure good vertical confinements of the modes,  the formed PPC patterns are air-suspended by undercutting the buried oxide layer. The PPC patterns are designed in a hexagonal lattice of air-holes with the lattice constant and radius of 450 nm and 110 nm, respectively. To introduce a defect in the PPC to form the cavity, the central four air-holes are shifted outwards~\cite{Zhang}. This point-shifted PPC cavity has an ultrasmall $V_{mode}$. Parts of air-holes surrounded the cavity defect are shrunk to shape an asymmetric mode distribution for high vertical coupling efficiency~\cite{Narimatsu}. 

The adopted AlGaAs NW was grown on a silicon substrate in a horizontal flow atmospheric pressure metalorganic vapor phase epitaxy system. The growth details and  characterizations of dimensions and crystallinity of NWs are discussed in the Supporting Information. In our experiment, a single NW is picked up from the vertically grown NW cluster and dropped onto the PPC region using a tungsten probe with the help of a high-resolution motorized stage. The effective coupling between the NW and cavity mode requires the NW to be placed on the cavity-area precisely, where the resonant mode locates. To realize that, an atomic force microscope (AFM) assisted transfer technique is exploited. We first examine the relative position of the NW to the cavity-area and then pull the NW accordingly to the cavity-area using the AFM probe, as indicated in Fig.~\ref{<device>}(a). Figure~\ref{<device>}(b) shows an  optical microscope image of a fabricated device. The AFM image is shown in the inset, showing the NW located at the middle of the cavity defect. The NW diameter is estimated around 133 nm from the AFM profile.  

\begin{figure}[th!]\centering
	\includegraphics[width=82.5mm]{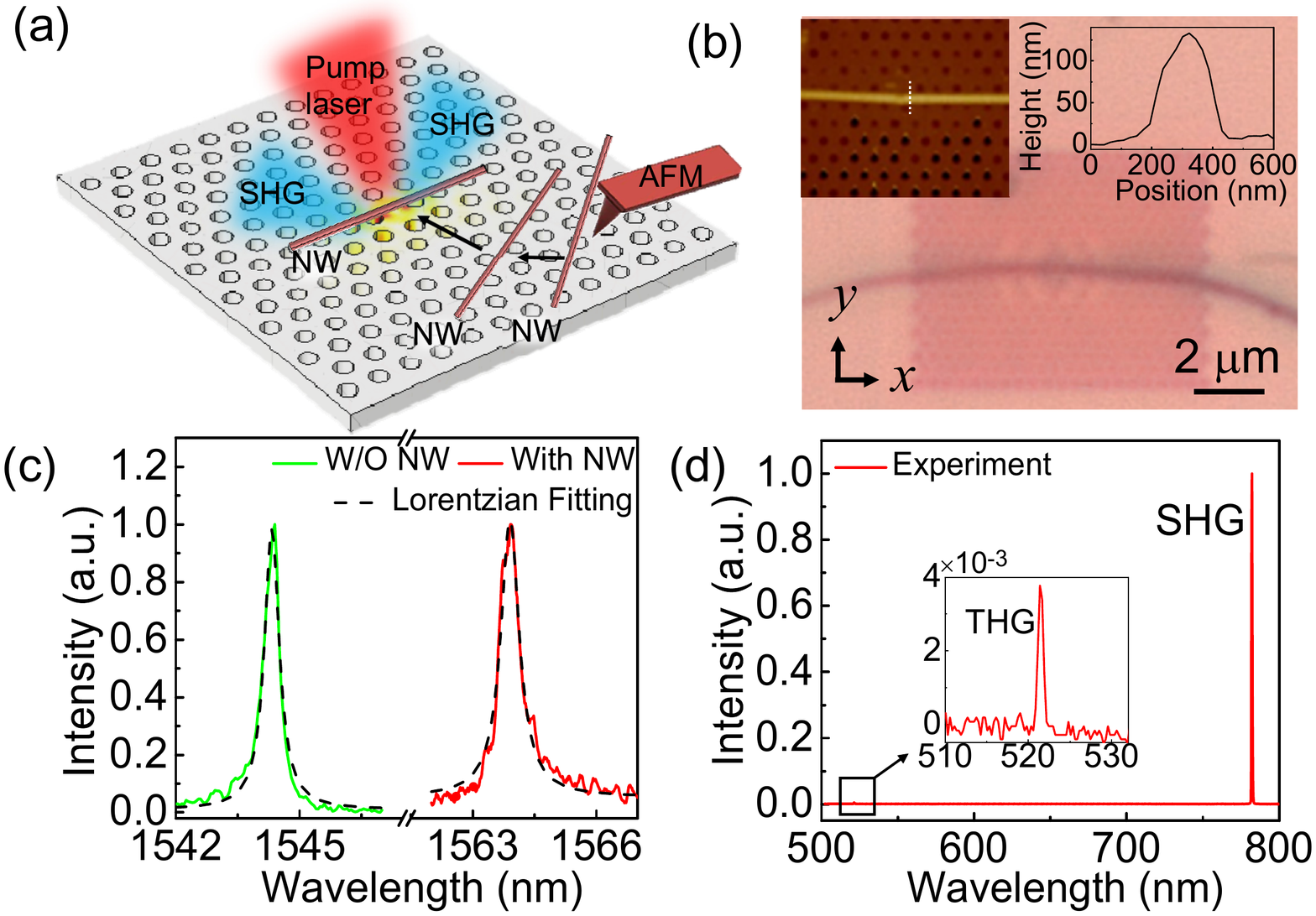}
	\caption{{\small (a) Schematics of the precise alignment of a NW onto a PPC cavity using AFM technique, and the operation of cavity-enhanced SHG including the vertically incident on-resonance pump and scattering SH signal from the NW. (b) Optical microscope image of the fabricated AlGaAs NW-PPC cavity. Insets show the AFM image as well as the height profile across the dashed white line. (c) Cavity's reflection spectra of the resonant mode before and after the integration of the AlGaAs NW.  Lorentzian fittings are included to estimate the $Q$ factors and resonant wavelengths. (d) Frequency-upconversion spectrum measured from the on-resonance pumped NW-PPC cavity. Inset plots the zoomed spectrum around 521 nm.} }
	\label{<device>} 
\end{figure}

\section{Results and discussions}

We characterize the resonant mode and SHG of the NW-PPC cavity using a vertically coupled cross-polarization microscope (see Supporting Information)~\cite{Gan2018}. Figure~\ref{<device>}(c) plots the resonant peak of a PPC cavity before and after the NW-integration. Fitted by Lorentzian functions, the resonant wavelengths and $Q$ factors are extracted. The AlGaAs NW works as a positive perturbation of the dielectric function around the resonant mode, hence, the resonant wavelength undergoes a red-shift from 1544.2 nm to 1564.2 nm. This 20 nm red-shift of the resonant wavelength indicates the effective coupling between the NW and the cavity. On the other hand, NW's high refractive index breaks the vertical total internal reflection of the silicon slab and induces extra scattering of the resonant mode. The $Q$ factor decreases from 3,500 to 2,800  after the NW integration. This slightly decreased $Q$ factor will not substantially degrade the cavity-enhanced SHG by considering that the intracavity electrical field is proportional to the $Q/V_{mode}$ factor. 

To pump the cavity-enhanced SHG from the NW-PPC cavity, we tune the laser wavelength as 1564.2 nm to resonantly excite the cavity mode. The frequency-upconversion signal is obtained by filtering out the pump laser with a short-pass dichroic mirror (see Supporting Information). Figure~\ref{<device>}(d) displays an acquired spectrum with a 2.3 mW pump power focused on the NW-PPC cavity. A strong peak is observed at the wavelength of 782.1 nm, corresponding to the half-wavelength of the pump laser. To verify that this peak originates from the second harmonic (SH) process, we examine its intensity dependence on the pump power, as shown in Fig.~\ref{<shg1>}(a). The pump power focused on the device is decreased from 2.3 mW to 0.05 mW.  The experimentally measured SH intensities are fitted well by a quadratic function of the pump power, a typical characteristic of SHG~\cite{Shen}. 

In the obtained frequency-upconversion spectrum, there is another weak peak at 521.4 nm, as shown in the zoomed plot of Fig.~\ref{<device>}(d). Confirmed from the pump power dependence, this peak corresponds to the third harmonic generation (THG) of the pump laser. Note the THG intensity is only $3.8\times10^{-3}$ times of the SHG intensity. Before the integration of NW, we also measure the frequency-upconversion signals from the bare silicon PPC cavity. By resonantly pumping the cavity mode with a laser at 1544.2 nm, only a weak THG peak is obtained, which could be attributed to silicon's intrinsic third-order nonlinearity. By comparing the THG signals obtained from the bare PPC cavity and NW-PPC cavity, their intensities are similar, which indicates that the THG peak of the NW-PPC cavity mainly comes from the silicon slab. There is no observable SHG from the bare silicon cavity due to silicon's inversion symmetry. Even if the inversion symmetry is broken at the silicon surface, the induced SHG should be two orders of magnitude weaker than the THG~\cite{Galli}.  Hence, the obtained strong SHG from the NW-PPC cavity mainly arises from the strong second-order nonlinearity of the AlGaAs NW. Calculated from the simulated resonant mode distribution of the NW-PPC cavity, the light power located in the NW is only around 0.87\% of that in the bulk silicon slab. Considering that AlGaAs and silicon have comparable third-order nonlinear coefficient, the THG from the AlGaAs NW has almost no contribution to the obtained THG. However, the SHG from the AlGaAs NW is remarkably strong, which could be considered as a new route to low-power nonlinear optical processes in silicon photonic devices. 

While the light power in the NW is no more than 1\% of that in the cavity mode, the generated SH signal is efficient enough. For instance, the SHG could be measured reliably even with a 50 $\mu$W CW laser power focused on the device (Fig.~\ref{<shg1>}(a)). And only 6\% of the CW pump light could be coupled into the cavity mode with the employed objective lens~\cite{Narimatsu,Gan2018}, i.e., the effective pump power is only $\sim$3 $\mu$W. In the previously reported NW SHG results,  pulsed light sources with a peak power around 100 W were typically used~\cite{Hu,Ren}. The low-power CW pumped SHG demonstrated here could be attributed to the strong cavity-enhancement. To verify that, we tune the CW laser wavelength across the resonant wavelength (at 1564.2 nm). The pump power illuminated on the device is maintained as 0.5 mW. The measured SH signals at various pump wavelengths are plotted in Fig.~\ref{<shg1>}(b), where the SH intensities are normalized by the SH intensity obtained with the on-resonance pump. When the laser wavelength is tuned away from the resonant mode, the SH signals decrease to an undetectable level. For a cavity mode, the densities of light power at different wavelengths are governed by a Lorentzian function $f_{Lorentzian}$, as indicated in Fig.~\ref{<device>}(c). In the SH process, the SHG intensity typically varies as the square of the input power of the fundamental wave~\cite{Shen}. Hence, the obtained SH spectrum should be determined by the squared Lorentzian function, as indicated by the fitting curve $(f_{Lorentzian})^2$ in Fig.~\ref{<shg1>}(b). 

\begin{figure}[th!]\centering
	\includegraphics[width=82.5mm]{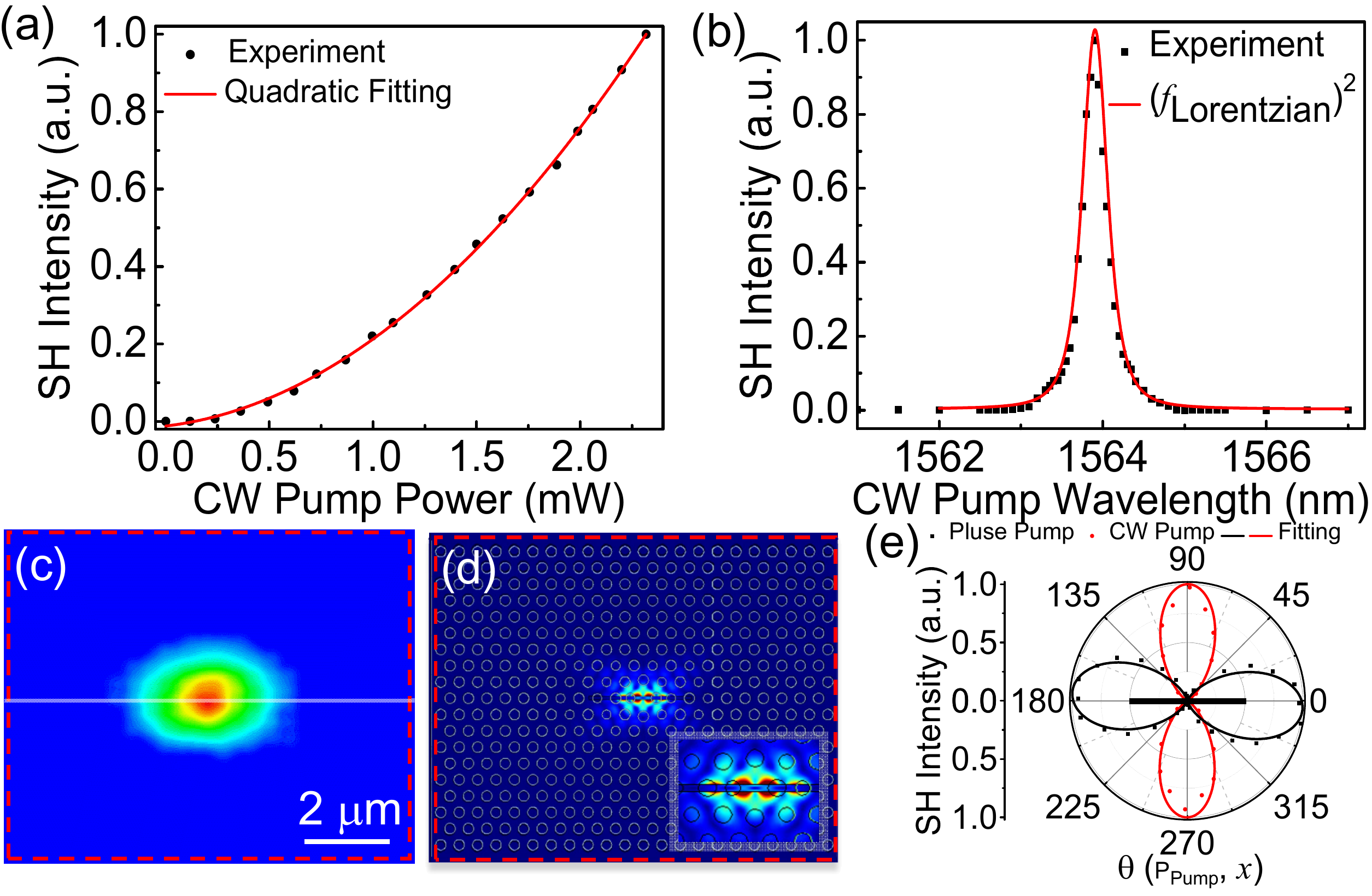}
	\caption{{\small (a) Power-dependence of the cavity-enhanced SHG fitted by a quadratic function of the pump power, which is measured after the objective lens. (b) SH intensities as a function of the excitation wavelengths. The fitting function $f_{Lorentzian}$ is the Lorentzian function used to fit the resonant mode in Fig.~\ref{<device>}(c).  (c) Spatial mapping of SHG from the NW-PPC cavity, where the PPC boundary and NW location are indicated by the dashed red box and solid white line.  (d) Simulated resonant mode of the NW-PPC cavity, and the mode distribution around the cavity area is zoomed in the inset. (e) Polarization-dependence of the SHG pumped by the on-resonance CW laser and off-resonance pulsed laser, which are fitted as functions of $\sin^6(\theta)$ and $\cos^4(\theta+0.27^\circ)$, respectively. Here, $\theta$ is the angle between the laser polarization and cavity's $x$-axis.}}
	\label{<shg1>} 
\end{figure}

The cavity-enhancement is further illustrated by implementing a SHG spatial mapping. The device is mounted on a two-dimensional piezo-actuated stage with a moving step of 0.1 $\mu$m, and the generated SH signals are monitored using a visible photomultiplier tube (PMT). Figure~\ref{<shg1>}(c) displays the measured result, where the red dashed box and white solid line indicate the PPC boundary and NW location, respectively. Figure~\ref{<shg1>}(d) shows the electrical field distribution of the resonant mode for the NW-PPC cavity at 1564.2 nm, and the inset displays the zoomed distribution around the cavity-area. The mode simulations are implemented using a finite element technique (COMSOL Multiphysics) with the structure parameters extracted from the fabricated NW-PPC cavity. The refractive indices of silicon and  AlGaAs NW around the wavelength of 1550 nm are chosen as 3.48 and 3.04, respectively~\cite{Aspnes}. Because the cavity-enhanced SHG in the NW is pumped by the resonant mode of the PPC cavity, this simulated distribution of the cavity mode could function as a reference to analyze the SHG spatial mapping shown in Fig.~\ref{<shg1>}(c).  The experiment results reveal that, only when the laser is focused around the cavity defect region, clear SH signals could be obtained due to the recirculatedly localized near-field of the resonant mode. Outside the mode distribution of the PPC cavity, the NW is only pumped by the vertically single-passed CW laser. The light-NW coupling is very weak for the nanoscale cross-section of a NW, which therefore makes it impossible to excite detectable SHG. The measured spatial profile of the SH signals has sizes around 2.5 $\mu$m and 1.8 $\mu$m along the $x$- and $y$-axis, respectively. Pumped by the cavity mode, the detected SH signals originate from the SH polarization generated in the  NW. Hence, the SHG spatial profile should be determined by the overlapped dimension between the NW and the cavity mode, as indicated by the zoomed distribution of the resonant mode shown in Fig.~\ref{<shg1>}(d). Considering that the NW aligns along the cavity's $x$-axis, the measured SHG profile in this direction is determined by the width of the cavity mode. Along the $y$-direction, SH signal radiates from SH dipoles distributing over a 133 nm scale. However, the employed microscope setup can only provide a spatial resolution of 1.8 $\mu$m (see Supporting Information), which limits the distinguishable SH profile along $y$-direction. 

Determined by crystal structure and $\chi^{(2)}$ tensor, the SHG in AlGaAs NWs exhibits varied strengths when the pump laser polarizes along different directions~\cite{Wang,He}.  Figure~\ref{<shg1>}(e) displays the measured NW's SH intensities pumped with varied  polarization directions, where $\theta$ is the angle between the laser polarization and $x$-axis of the PPC cavity. These SHG polarization-dependences are studied using different pump lasers, including the on-resonance CW laser and an off-resonance pulsed laser (a picosecond pulsed laser at 1560 nm with a repetition rate of 18.5 MHz and a pulse width of 8.8 ps). Such a picosecond laser pump configurations were widely exploited in previously reported works~\cite{Wang,He}. The obtained result with the off-resonance pulsed pump is fitted by a function of $\cos^4(\theta+0.27^\circ)$, which is coincident with the polarization-dependence determined by AlGaAs NW's $\chi^{(2)}$ tensor~\cite{Wang,He}. When the laser polarization is perpendicular to NW's longitudinal axis, the SH intensity approaches to zero. 

Pumped by the on-resonance CW laser (at 1564.2 nm), the SHG polarization-dependence has a $\sim90^\circ$ difference from that obtained with the pulsed laser, as shown in the red dots of Fig.~\ref{<shg1>}(e). It could be attributed to the cavity-enhanced interaction between the NW and cavity's near-field. When the on-resonance CW laser is focused on the NW-PPC cavity, part of the focused light will couple into the cavity mode to excite a resonantly localized near-field around the NW, which is strong enough to pump NW's SHG. The electrical field distribution of the resonant mode is only determined by the physical structure of the NW-PPC cavity. Hence, the coupling between the mode field and NW's nonlinear coefficient tensor $\chi^{(2)}$ would not be modified by the laser polarization. However, for the on-resonance CW laser with different polarizations, the coupling efficiencies into the resonant mode are changed significantly~\cite{Narimatsu}, which gives rise to varied strengths of the cavity mode for generating different SH intensities. Determined by the far-field radiation pattern of the employed resonant mode, only the on-resonance laser polarized along the $y$-axis could strongly couple with it. Hence, for an incident light with an electrical field of $E_P$, the electrical field coupled into the resonant mode is ${\eta}E_P\sin(\theta)$, where $\eta$ is the maximum coupling efficiency. With that, considering the three-wave mixing process, the electrical field to generate SH signal is proportional to $({\eta}E_P\sin(\theta))^2$, and the SH power has a  $\sin^4(\theta)$ dependence on the laser polarization direction. Considering the cross-polarization of the experimental setup, the vertically scattered SHG after the HWP is then projected onto the output polarization direction, and the finally collected SHG thus has a function of $\sin^6(\theta)$, as indicated by the fitting line shown in Fig.~\ref{<shg1>}(e).

\begin{figure}[th!]\centering
	\includegraphics[width=82.5mm]{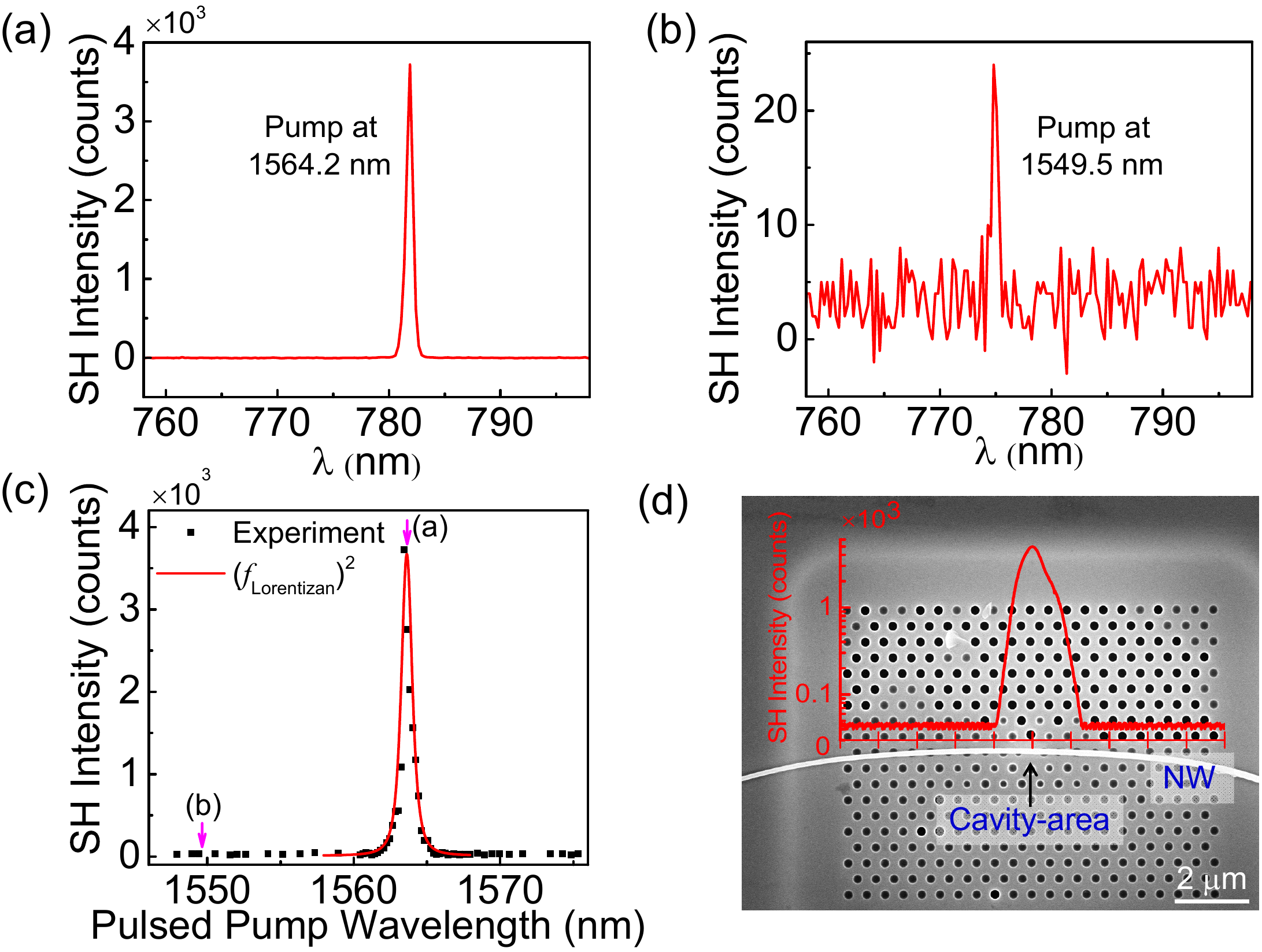}
	\caption{{\small {(a, b) SHG spectra obtained with pulsed pumps at 1564.2 nm  (a)  and 1549.5 nm (b), which are acquired with the same pump power and same integration time of the detector. (c) SH intensities as a function of the pump wavelengths. (d) SH intensities obtained by focusing the on-resonance pulsed laser on different locations of the NW, which is positioned on the device's SEM image to hint the signal-collection positions.}}}
	\label{<ps>} 
\end{figure}

The results shown in Figs.~\ref{<shg1>}(b) and~\ref{<shg1>}(c) indicate the capability of low-power CW pumped SHG from a NW with the significant cavity-enhancement. However, when the laser wavelength is tuned away from the cavity's resonant wavelength (Fig.~\ref{<shg1>}(b)), or the laser is focused on the NW located outside the cavity-area (Fig.~\ref{<shg1>}(c)), there is no any detectable SH signals for the weak light-NW interaction. This zero SH signal from the bare NW  makes it impossible to do the comparison with the cavity-enhanced SHGs for evaluating the enhancement factor. To estimate this enhancement factor, we further study the pump wavelength and spatial location dependences of NW's SHG by pumping the NW-PPC cavity with a pulsed laser (see Supporting Information), which could yield detectable SH signals from NWs uncoupled with the cavity. 

In the measurements, we first focus the pulsed laser over the cavity-area and tune its wavelength across the cavity mode. The pump power is fixed. Figures~\ref{<ps>}(a) and (b) display the SHG spectra acquired when the pump wavelengths are 1564.2 nm  and 1549.5 nm, corresponding to the on- and off-resonance pumps, respectively. With the same integration time, the peak photon numbers recorded by the spectrometer for the two cases are 3722 counts and 25 counts, giving rise to a cavity-enhancement factor of 149.  This enhancement is represented further by plotting the  peak SH photon numbers  obtained with a broad pump wavelength range, as shown in Fig.~\ref{<ps>}(c). Similar as that shown in Fig.~\ref{<shg1>}(b), with the significantly strong SHG pumped at 1564.2 nm, the wavelength dependence is well  fitted by $(f_{Lorentzian})^2$. Next, by fixing the laser wavelength at 1564.2 nm, we focus the pulsed laser on different locations along the NW. The corresponding peak SH photon numbers  are plotted in Fig.~\ref{<ps>}(d), where the scanning electron microscope (SEM) image of the device is displayed as well to hint the signal-collection positions.  This spatial dependence is consistent with that obtained with the CW pump, showing remarkable SH signal (photon counts of 5000) from the NW located within the cavity-area.  When the NW is outside the cavity-area, the pulsed laser pumps the bare NW without cavity-enhancement and yields a SH photon counts only around 35. This result indicates an enhancement factor about 143, which is close to that evaluated from the wavelength-dependent SHG. 

\begin{figure}[th!]\centering
	\includegraphics[width=82.5mm]{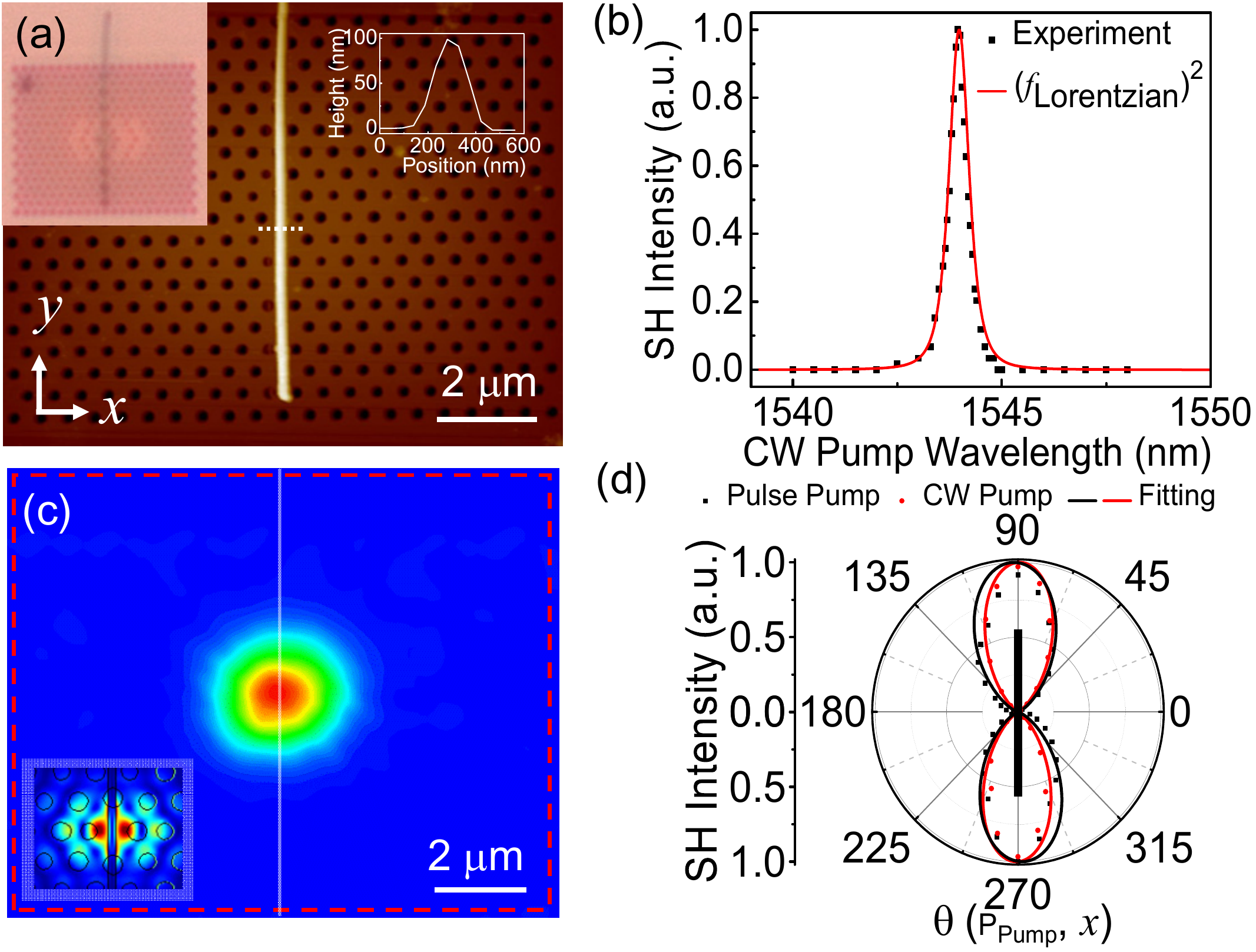}
	\caption{{\small (a) AFM image of another AlGaAs NW-PC cavity, where the axis of NW is aligned along y-axis. Left inset displays optical microscope image of the sample; Right inset shows the height profile along the dashed white line in the AFM image. (b)  SH intensity as a function of the excitation wavelength. (c) Spatial mapping of SHG from the NW-PPC cavity, where the PPC boundary and NW location are indicated by the dashed red box and solid white line. Inset shows the simulated electrical field distribution of the cavity's mode around the cavity-area. (d) Polarization-dependence of the SHG pumped by the on-resonance CW laser and the off-resonance pulsed laser, which are fitted as functions of $\sin^6(\theta)$ and $\sin^4(\theta+0.6^\circ)$, respectively.  }}
	\label{<shg2>} 
\end{figure}

Considering NW's one-dimensional structure and cavity-mode's two-dimensional distribution, the cavity-enhanced SHG would be valid for NWs orientating along other directions, given that the effective coupling between the NW and the resonant mode is maintained. We fabricate another device with a AlGaAs NW parallel to the cavity's $y$-axis. The AFM and optical microscope images shown in Fig.~\ref{<shg2>}(a) indicate that  the NW locates on the cavity center precisely. The measured NW diameter is about 98.8 nm, and the length is estimated as 7 $\mu$m. In this device, the integration of NW  induces a red-shift of the resonant wavelength by 15 nm and degrades the $Q$ factor from 2,300 to 1,800. From the electromagnetic perturbation theory of cavity mode~\cite{Joannopoulos}, this less shifted resonant wavelength indicates weaker light-NW coupling than that achieved in the previous device (Fig.~\ref{<device>}(b)). Pumped by an on-resonance CW laser with hundreds of microwatts focused on the device, strong SHG is obtained as well. 

The cavity-enhancement of this device is verified by tuning the CW laser wavelength across the resonant wavelength, as shown in Fig.~\ref{<shg2>}(b). The experimental results are fitted very well by the squared Lorentzian lineshape of the fundamental resonance. Figure~\ref{<shg2>}(c) displays the spatial mapping image of the device's SHG, and the inset shows the electrical field distribution of the resonant mode for the NW-PPC cavity around the cavity-area. Different from that shown in Fig.~\ref{<shg1>}(c), it has a circular profile with a diameter around 1.8 $\mu$m. As mentioned above, the NW is considered as an component of SH emitters with a dimension overlapping with the cavity mode. Hence, the SH radiation along the $y$-direction has a scale of 1.8 $\mu$m determined by the vertical cavity mode distribution. Along the $x$-axis, the diffraction limit of the microscope gives rise to a 1.8 $\mu$m lateral size of the SHG mapping, though the SH emitters only locate in a diameter of 98.8 nm. Figure{~\ref{<shg2>}(d) plots the SHG variations with respect to the pump lasers' polarization directions. Pumped by the on-resonance CW laser, the SHG polarization-dependence has a function of $\sin^6(\theta)$, which is similar as the result presented in the previous device (Red curve in Fig. 2(e)). This could be attributed to the near-field enhanced light-NW  interaction and polarization-dependent far-field coupling of the cavity mode. For the off-resonance pulsed pump, the measured polarization-dependent SHG is fitted by a function of $\sin^4(\theta+0.6^\circ)$, which agrees with the SHG process determined by NW's crystal structure. Also, this result indicates that the NW has a rotating angle of $0.6^\circ$ from the $y$-axis. 
	
\section{Conclusion}
In conclusion, we have demonstrated effectively enhanced SHG in NWs by integrating them with silicon PPC cavities. Assisting by the extremely localized resonant mode and NW's high second-order optical nonlinearity, it is possible to realize efficient SHGs with a CW laser, and the pump power could be greatly reduced less than 10 $\mu$W. Comparing with SHGs pumped by off-resonance and on-resonance pulsed lasers, the on-resonance pumped SHG is enhanced by factors around 150. In the NW-PPC cavity, NWs' SHG is more than two-orders of magnitude stronger than the THG in the silicon slab, though they only couple with less than 1\% of the cavity mode. Semiconductor NWs have been integrated onto photonic chips to serve as waveguides~\cite{Park}, couplers~\cite{Chen}, switchinges~\cite{Piccione2}. Here, with the assistance of a silicon PPC cavity, the power requirement of SHG in NWs is reduced significantly. It will promote the NW-based on-chip nonlinear optical processes for coherent light sources, entangled photon-pairs and signal processing.
	
\begin{acknowledgement}

Financial support was provided by the NSFC (Grant Nos. 61522507, 61775183, 11634010), the Key Research and Development Program (Grant No. 2017YFA0303800), the Key Research and Development Program in Shaanxi Province of China (2017KJXX-12),  the Academy of Finland (Grant Nos. 276376, 284548, 295777, 304666, 312297, 312551, 284529, and 314810), TEKES (OPEC), the European UnionÕs Seventh Framework Program (Grant No.631
610), Aalto Centre of Quantum Engineering, China Scholarship Council, Walter Ahlstrom Foundation from Aalto University Doctoral school, and the provision of technical facilities of the Micronova, Nanofabrication Centre of Aalto University, and Analytical \& Testing Center of NPU.
\end{acknowledgement}


\section{Supporting Information}

\section{Table of contents:}

1. Growth recipes of the AlGaAs nanowires

\noindent 2. Characterizations of nanowires' crystal structures

\noindent3. Experimental arrangements

\noindent4. Reference	
	
\section{1. Growth recipes of the AlGaAs nanowires}

AlGaAs nanowires (NWs) were synthesized on Si (111) substrates using horizontal flow atmospheric pressure metalorganic vapor phase epitaxy (MOVPE) system. Trimethylaluminum (TMAl), trimethylgallium (TMGa), tertiarybutylarsene (TBAs) were used as precursors. Hydrogen was used as a carrier gas with the total reactor gas flow rate of 5 sccm. The substrates were first prepared inside an ultrasonic bath by immersing them in acetone and in isopropanol for 2 minutes each, followed by a 2 minutes rinse in deionized water. Gold nanoparticles with diameters around 40 nm in a colloidal solution (BBI International, UK) were used as catalysts for the vapor-liquid-solid (VLS) growth. Poly-L-Lysine (PLL) solution was applied to the substrate for 2 minutes for better nanoparticles adhesion, followed by 2 minutes deposition of gold nanoparticles. Prior to the growth, the substrates were annealed in situ at 650$^{\circ}$C for 10 minutes under hydrogen flow to desorb surface contaminants. The growth step was initiated by switching on the TMAl, TMGa and TBAs sources simultaneously for 5 minutes at a fixed growth temperature of 450$^{\circ}$C. The nominal V/III ratio during the growth was $\sim$11, and the TMAl, TMGa and TBAs flows were 20, 7 and 75 sccm, respectively. After the growth, only the TBAs flow was kept on during the reactor cooled down to 250 $^{\circ}$C. The temperatures reported in this work are thermocouple readings of the lamp-heated graphite susceptor, which are somewhat higher than the real substrate surface temperature.	
	
\section{2. Characterizations of nanowires' crystal structures}
To examine the crystal structures of the employed AlGaAs NWs, high-resolution transmission electron microscopy (TEM) measurements were implemented using a FEI Talos F200X operated at 200 kV. The grown NWs were transferred from the growth substrate onto a TEM copper grid using a tungsten probe. The measurement results are shown in Figure S1. The NW diameters varied from 50 nm to 150 nm were verified in the low-magnification mode, as shown in Figures S1(a) and (b). The high-resolution image and the select area electron diffraction pattern shown in Figures S1(c) and (d) indicate the predominant Wurtzite (WZ) crystal structure with a small concentration of crystal defects. There is no core/shell structure. To evaluate elemental compositions of the AlGaAs NWs, the scanning TEM integrated energy-dispersive X-ray spectroscopy was employed. The plane scanning results of the Al and Ga elements of the sample shown in Figure S1(e) were shown in Figures S1(f) and (g), respectively, which indicate the elemental uniformity of the grown NW. Figure S1(h) displays the Al and Ga components (wt\%) along the dashed line shown in Figure S1(e), showing the Al composition around 20\%.
	
\begin{figure}[th!]\centering
	\includegraphics[width=82.5mm]{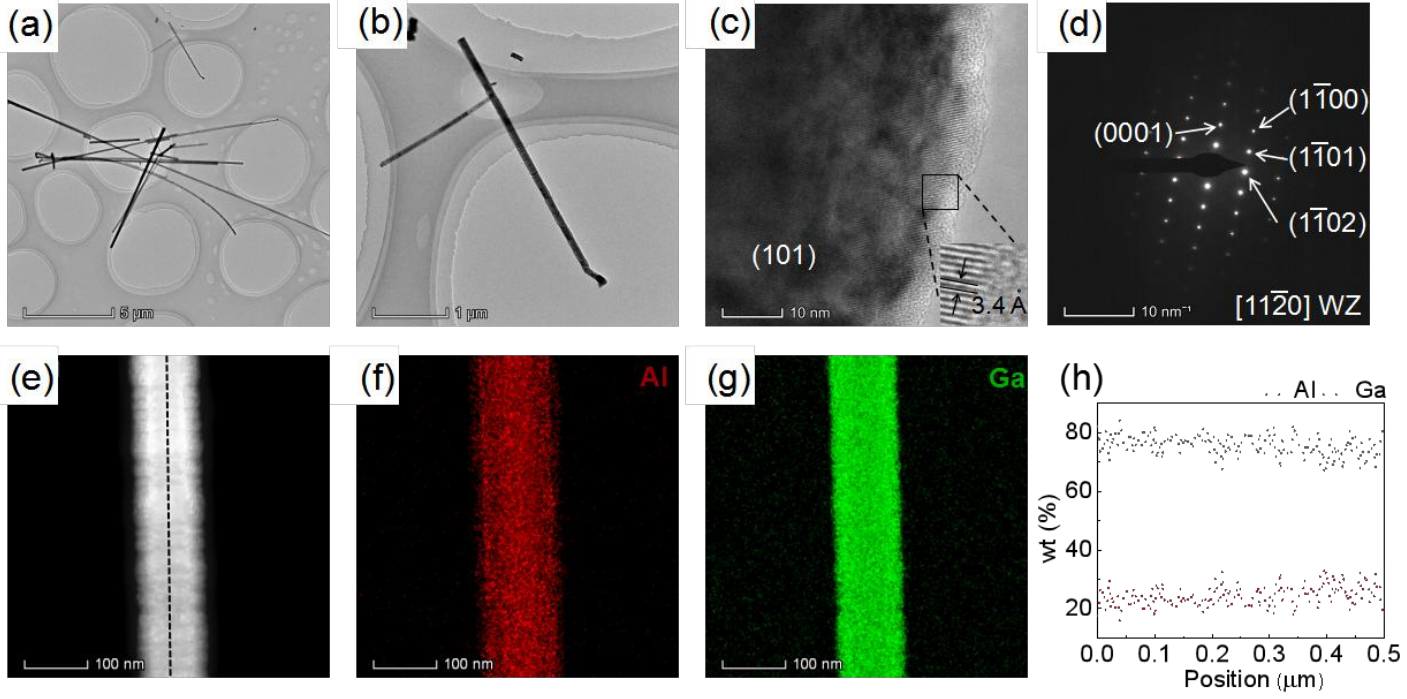}
\end{figure}
\begin{spacing}{1.2}
\small \noindent Figure S1: TEM characterization of the AlGaAs NW. (a, b) TEM images with different low-magnifications. (c) High-resolution image showing the predominant WZ structure with a small concentration of crystal defects. (d) Select area electron diffraction pattern demonstrating the WZ crystal structure. (f, g) Al and Ga components mapping of the NW shown in (e). (h) Al and Ga components (wt\%) distribution along the black dashed line in (e).
\end{spacing}

\section{3. Experimental arrangements}
We experimentally measured the resonant modes and second harmonic generation (SHG) signal from the NW-PPC cavity using a vertically coupled cross-polarization microscope,$^{1}$ as schematically shown in Figure S2. 
To couple the excitation laser into the microscope setup, a dichroic mirror was employed to reflect it vertically into the objective lens. This dichroic mirror (DMSP1000, Thorlabs Inc.) is a short-pass one with a cutoff wavelength of 1000 nm. Light beams with wavelengths longer (shorter) than 1000 nm will be reflected (transmit) by this dichroic mirror. Hence, if there is any second harmonic (SH) signal existing in the pump laser, it will transmit through the dichroic mirror and will not couple into the objective lens. An additional visible-range white light source was coupled into the microscope setup by transmitting it through the dichroic mirror, which will illuminate the sample via the objective lens. The reflected withe light from the sample will be collected and recorded by the visible CCD to monitor the position of the NW-PPC cavity. A polarized beam splitter (PBS) was employed in the microscope system to achieve orthogonally polarized excitation laser and collection signal, allowing for distinguishing the reflection of the cavity mode with a high signal to noise ratio. A half wave plate (HWP) was inserted between the PBS and the objective lens to control the direction of laser polarization with respect to the axis of the NW-PPC cavity. 
The objective lens of the microscope is a near-infrared anti-reflective one with a 50× magnification and a numerical aperture of 0.42. Here, considering the incident pump laser fills the input aperture of the objective lens, the focused beam diameter of the pump laser could be calculated by
\begin{equation}
d\approx\frac{\lambda}{2 \times NA} 
\end{equation}	
where $\lambda$ is the pump wavelength around 1550 nm, and \textit{NA}=0.42 is the numerical aperture of the objective lens. The focused pump beam diameter was estimated around 1.8 $\mu$m.

To examine the resonant peak of the PPC cavity, a narrowband tunable telecom-band CW laser (Yenista, T100S-HP/CL) was employed as the excitation source. Its reflection signal from the cavity was vertically collected by the objective lens and finally reflected by another short-pass dichroic mirror (DMSP1000, Thorlabs Inc.) into a telecom-band photodiode after passing through the PBS and HWP. By tuning the laser wavelength with a step of 0.005 nm, the cavity’s reflection spectra were obtained, as shown in Figure 1(c) of the maintext.

\begin{figure}[th!]\centering
	\includegraphics[width=82.5mm]{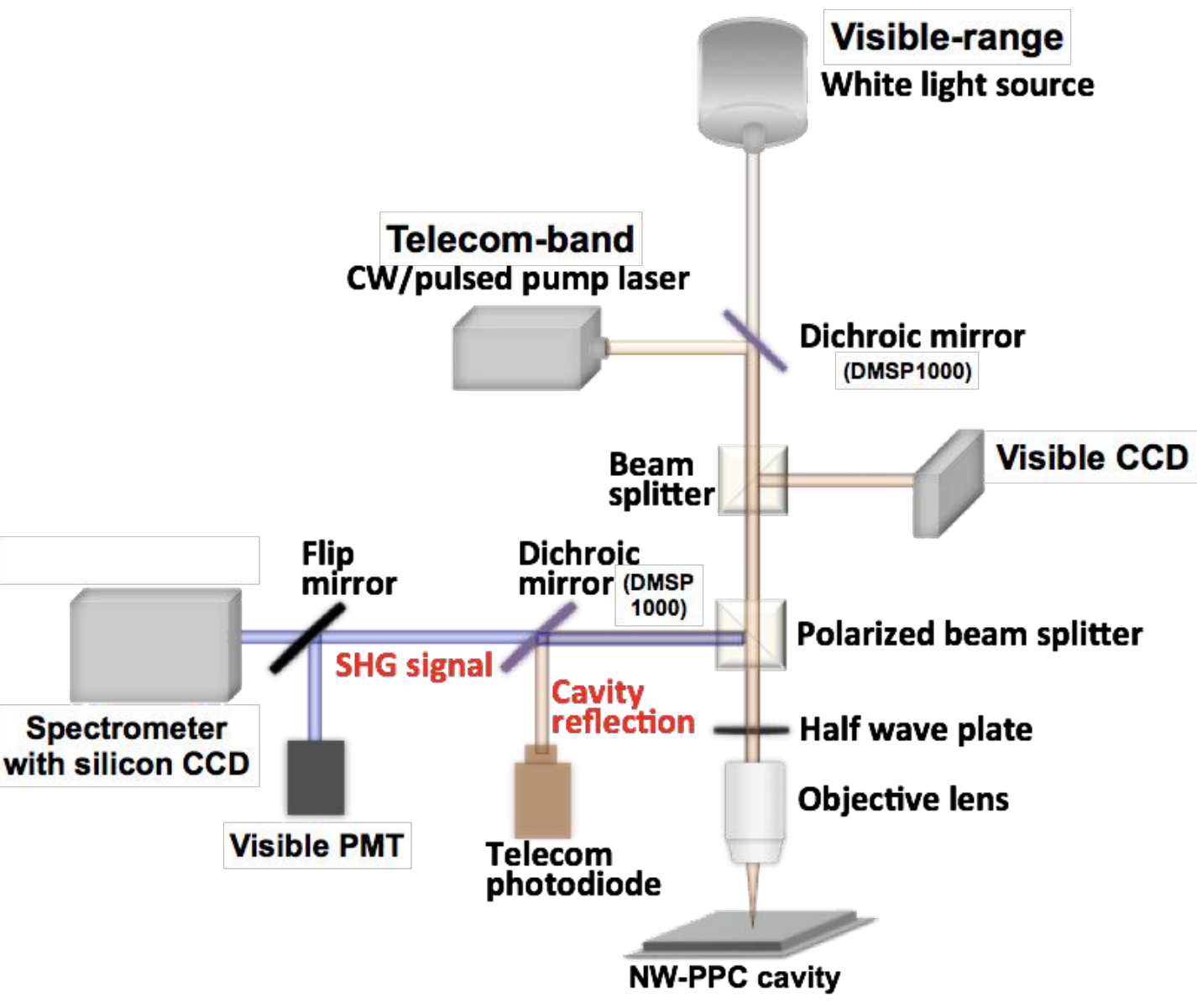}
\end{figure}
\begin{spacing}{1.2}
	\begin{center} 
		\small Figure S2: Schematic diagram of the measurement setup.
	\end{center}
\end{spacing}

For the measurements of SHG from the NW-PPC cavity, the telecom-band CW or pulsed laser was tuned on-resonance (at the wavelength of 1564.2 nm) to excite the cavity resonant mode. The frequency conversion signals scattered from the PPC cavity were collected by the objective lens, which passed through the short-pass dichroic mirror and were examined by a 0.5 m spectrometer (SP 2500i, Princeton Instruments Inc.) mounted with a cooled silicon camera (PIXIS 100 BX, Princeton Instruments Inc.). The SHG signal could be redirected into a visible photomultiplier tube (PMT) as well by a flipping mirror for measuring the SHG spatial mapping. As indicated by the Figure 1(d) of the maintext, SHG and third harmonic generation (THG) are obtained. 

The reflected pump laser from the PPC cavity is also collected by the objective lens. After passing through the HWP and PBS, the collected pump laser is reflected by the short-pass dichroic mirror. Hence, the dichroic mirror functions as a filter of the pump laser. However, even the dichroic mirror could filter the pump laser to a ratio of 0.7\% (data from the Thorlabs Inc.), the residual pump laser is still much stronger than the SHG signal. Fortunately, both of the visible PMT and the spectrometer with silicon camera will not sense the telecom-band pump laser, which ensures the signals recorded by the PMT and spectrometer are SHG or THG signals. 

Finally, to evaluate the cavity enhancement factor over the SHG, the CW pump laser was switched into a pulsed picosecond laser (PriTel FFL-20MHz), whose wavelength could be tuned between 1530 and 1570 nm. The pulsed laser was also tuned across the resonant mode of the NW-PPC cavity. Here, because of the high peak intensity of the pulsed laser, even the wavelength of the pulsed laser is off-resonance from the resonant mode of the NW-PPC cavity, there is still detectable SHG from the NW. Hence, by pumping the NW-PPC cavity with the pulsed laser at different wavelengths, the enhancement factor could be extracted by comparing the NW’s SHG intensities when the pulsed laser was no-resonance and off-resonance with the cavity, as shown in Figures 3(a)-(c) of the maintext. In addition, the enhancement factor could be evaluated by comparing the SHGs obtained when the pulsed laser is focused on the NW inside and outside the cavity region, while the wavelength of the pulsed laser is maintained on-resonance with the cavity mode. When the laser is focused on the NW inside the cavity region, its SHG will be enhanced by the cavity mode significantly. For the NW outside the cavity region, the focused laser will only pump on the bare NW, and the SHG signal would be very small due to the weak light-NW interaction. As shown in Figure 3(d) of the maintext, the SH photon counts obtained from the NW inside and outside the cavity region are 5000 and 35, respectively, which indicates an enhancement factor of 143.

\section{4. Reference}
(1) X. Gan, C. Zhao, S. Hu, T. Wang, Y. Song, J. Li, Q. Zhao, W. Jie, and J. Zhao, Light : Science $\&$ Applications 7, 17126 (2018).


\begin{thebibliography}{37}
	\bibitem{Agarwal} R. Agarwal, C. J. Barrelet, and C. M. Lieber, \textit{Nano Lett.} \textbf{5},  917--920 (2005).
	
	\bibitem{Hayden}O. Hayden, R. Agarwal, and C. M. Lieber, \textit{Nat. Mater.} \textbf{5}, 352 (2006).
	
	\bibitem{Fang}H. Fang, W. Hu, P. Wang, N. Guo, W. Luo, D. Zheng, F. Gong, M. Luo, H. Tian, and X. Zhang, \textit{Nano Lett.} \textbf{16}, 6416--6424 (2016).
	
	\bibitem{Piccione}B. Piccione, L. K. van Vugt, and R. Agarwal, \textit{Nano Lett.} \textbf{10}, 2251--2256 (2010).
	
	\bibitem{Johnson}J. C. Johnson, H. Yan, R. D. Schaller, P. B. Petersen, P. Yang, and R. J. Saykally, \textit{Nano Lett.} \textbf{2}, 279-283 (2002).
	
	\bibitem{Liu}W. Liu, K. Wang, Z. Liu, G. Shen, and P. Lu, \textit{Nano Lett.} \textbf{13}, 4224--4229 (2013).
	
	\bibitem{Barzda}V. Barzda, R. Cisek, T. Spencer, U. Philipose, H. Ruda, and A. Shik, \textit{Appl. Phys. Lett.} 92, 113111 (2008).
	
	\bibitem{Hu}H. Hu, K. Wang, H. Long, W. Liu, B. Wang, and P. Lu, \textit{Nano Lett.} \textbf{15}, 3351--3357 (2015).
	
	\bibitem{Yu}H. Yu, W. Fang, X. Wu, X. Lin, L. Tong, W. Liu, A. Wang, and Y. R. Shen, \textit{Nano Lett.} \textbf{14}, 3487--3490 (2014).
	
	\bibitem{Xin}C. Xin, S. Yu, Q. Bao, X. Wu, B. Chen, Y. Wang, Y. Xu, Z. Yang, and L. Tong, \textit{Nano Lett.} \textbf{16}, 4807--4810 (2016).
	
	\bibitem{Chen1}R. Chen, S. Crankshaw, T. Tran, L. C. Chuang, M. Moewe, and C. Chang-Hasnain, \textit{Appl. Phys. Lett.} \textbf{96}, 051110 (2010).
	
	\bibitem{Long}J. Long, B. Simpkins, D. Rowenhorst, and P. Pehrsson, \textit{Nano Lett.} \textbf{7}, 831--836 (2007).
	
	\bibitem{Zhuo}G. Y. Zhuo, K. J. Hsu, T. Y. Su, N.-H. Huang, Y. F. Chen, and S. W. Chu, \textit{J. Appl. Phys.} \textbf{111}, 103112 (2012).
	
	\bibitem{Sanatinia}R. Sanatinia, M. Swillo, and S. Anand, \textit{Nano Lett.} \textbf{12}, 820--826 (2012).
	
	\bibitem{Han}X. Han, K. Wang, H. Long, H. Hu, J. Chen, B. Wang, and P. Lu, \textit{ACS Photonics} \textbf{3}, 1308--1314 (2016).
	
	\bibitem{Jassim}N. M. Jassim, K. Wang, X. Han, H. Long, B. Wang, and P. Lu, \textit{Opt. Mater.} \textbf{64}, 257--261 (2017).
	
	\bibitem{Neeman}L. Neeman, R. Ben-Zvi, K. Rechav, R. Popovitz-Biro, D. Oron, and E. Joselevich, \textit{Nano Lett.} \textbf{17}, 842--850 (2017).
	
	\bibitem{Ren}M. L. Ren, R. Agarwal, W. Liu, and R. Agarwal, \textit{Nano Lett.} \textbf{15}, 7341--7346 (2015).
	
	\bibitem{Pimenta}  A. C. S. Pimenta, D. C. Teles Ferreira, D. B. Roa, M. V. B. Moreira, A. G. de Oliveira, J. C. Gonz{\'a}alez, M. De Giorgi, D. Sanvitto, and F. M. Matinaga, \textit{The Journal of Physical Chemistry C} \textbf{120}, 17046--17051 (2016).
	
	\bibitem{Timofeeva} M. Timofeeva, A. Bouravleuv, G. Cirlin, I. Shtrom, I. Soshnikov, M. Reig Escal{\'e}, A. Sergeyev, and R. Grange, \textit{Nano Lett.} \textbf{16}, 6290--6297 (2016).  
	
	\bibitem{Liu2}C.-W. Liu, S.-J. Chang, C.-H. Hsiao, R.-J. Huang, Y.-S. Lin, M.-C. Su, P.-H. Wang, and K.-Y. Lo, \textit{IEEE Photonics Technol. Lett.} \textbf{26}, 789--792 (2014).
	
	\bibitem{Nakayama}Y. Nakayama, P. J. Pauzauskie, A. Radenovic, R. M. Onorato, R. J. Saykally, J. Liphardt, and P. Yang, \textit{Nature} \textbf{447}, 1098 (2007).
	
	\bibitem{Grinblat}G. Grinblat, M. Rahmani, E. Cortés, M. Caldarola, D. Comedi, S. A. Maier, and A. V. Bragas, \textit{Nano Lett.} \textbf{14}, 6660-6665 (2014).
	
	\bibitem{Casadei}A. Casadei, E. F. Pecora, J. Trevino, C. Forestiere, D. Rüffer, E. Russo-Averchi, F. Matteini, G. Tutuncuoglu, M. Heiss, A. Fontcuberta i Morral, and L. Dal Negro, \textit{ Nano Lett.} \textbf{14}, 2271--2278 (2014).
	
	\bibitem{Richter}J. Richter, A. Steinbrück, T. Pertsch, A. Tünnermann, and R. Grange, \textit{Plasmonics} \textbf{8}, 115--120 (2013).
	
	\bibitem{Sergeyev}A. Sergeyev, R. Geiss, A. S. Solntsev, A. A. Sukhorukov, F. Schrempel, T. Pertsch, and R. Grange, \textit{ACS Photonics} \textbf{2}, 687--691 (2015).
	
	\bibitem{Englund}  D. Englund, D. Fattal, E. Waks, G. Solomon, B. Zhang, T. Nakaoka, Y. Arakawa, Y. Yamamoto, and J. Vu{\v{c}}kovi{\'c}, \textit{Phys. Rev. Lett.} \textbf{95}, 013904 (2005).
	
	\bibitem{Prieto}I. Prieto, J. Llorens, L. E. Munoz-Cam{\'u}nez, A. Taboada, J. Canet-Ferrer, J. M. Ripalda, C. Robles, G. Mu{\~n}oz-Matutano, J. Mart{\'\i}nez-Pastor, and P. A. Postigo, \textit{Optica} \textbf{2}, 66--69 (2015).
	
	\bibitem{Yokoo}A. Yokoo, M. Takiguchi, M. D. Birowosuto, K. Tateno, G. Zhang, E. Kuramochi, A. Shinya, H. Taniyama, and M. Notomi, \textit{ACS Photonics} \textbf{4}, 355--362 (2017).
	\bibitem{Zhang}Z. Zhang and M. Qiu, \textit{Opt. Express} \textbf{12}, 3988--3995 (2005).
	
	\bibitem{Narimatsu}M. Narimatsu, S. Kita, H. Abe, and T. Baba, \textit{Appl. Phys. Lett.} 100, 121117 (2012).
	
	\bibitem{Gan2018}X. Gan, C. Zhao, S. Hu, T. Wang, Y. Song, J. Li, Q. Zhao, W. Jie, and J. Zhao,  \textit{Lig. Sci. Appl.} 7, 17126 (2018). 
	
	\bibitem{Shen}Shen, Y. R. The principles of nonlinear optics; Wiley-Interscience: New York, 2003.
	
	\bibitem{Galli}M. Galli, D. Gerace, K. Welna, T. F. Krauss, L. O'Faolain, G. Guizzetti, and L. C. Andreani, \textit{Opt. Express} \textbf{18}, 26613-26624 (2010).
	
	\bibitem{Aspnes} D.E. Aspnes, S. M. Kelso, R. A. Logan, and R. Bhat, \textit{J. Appl. Phys.} \textbf{60}, 754--767 (1986).
	
	\bibitem{Wang}J. Wang, Y. Yu, Y. M. Wei, S. F. Liu, J. Li, Z. K. Zhou, Z. C. Niu, S. Y. Yu, and X. H. Wang, \textit{Sci. Rep.} \textbf{7}, 2166(2017).
	
	\bibitem{He}H. He, X. Zhang, X. Yan, L. Huang, C. Gu, M. l. Hu, X. Zhang, X. M. Ren, and C. Wang, \textit{Appl. Phys. Lett.} \textbf{103}, 143110 (2013).
	
	\bibitem{Joannopoulos}J. D. Joannopoulos, S. G. Johnson, J. N. Winn, and R. D. Meade, Photonic crystals: molding the flow of light (Princeton university press, Princeton, (2011) p. 35. 
	
	\bibitem{Park}H. G. Park, C. J. Barrelet, Y. Wu, B. Tian, F. Qian, and C. M. Lieber, \textit{Nat. Photonics} \textbf{2}, 622 (2008).
	
	\bibitem{Chen}B. Chen, H. Wu, C. Xin, D. Dai, and L. Tong, \textit{Nat. Commun.} \textbf{8}, 20 (2017).
	
	\bibitem{Piccione2}B. Piccione, C. H. Cho, L. K. Van Vugt, and R. Agarwal, \textit{Nat. Nanotechnol.} \textbf{7}, 640 (2012).
	
\end{thebibliography}
\end{document}